\documentclass[10pt,twocolumn]{article}

\usepackage{isqcmc2023}
\usepackage{graphicx,url}
\usepackage{amsmath}

\usepackage[utf8]{inputenc}

\title{Towards the Intuitive Understanding of Quantum World: Sonification of Rabi Oscillations, Wigner functions, and Quantum Simulators}

\author{Reiko Yamada\inst{1} \and
Eloy Piñol\inst{1} \and
Samuele Grandi\inst{1} \and
Jakub Zakrzewski\inst{2} \inst{3} \and
Maciej Lewenstein\inst{1} \inst{4}}

\address{ICFO -- Institut of Photonic Science, Barcelona Institute of Science and Technology\\
         Av. C.F. Gauss, 3 - 08860 - Castelldefels - Barcelona - Spain
         \nextinstitute
         Institute of Theoretical Physics -- Jagiellonian University in Krak\'ow \\ \L ojasiewicza 11, 30-348 - Krak\'ow - Poland
         \nextinstitute
         Mark Kac Complex Systems Research Center -- Jagiellonian University \\ \L ojasiewicza 11, 30-348 - Krak\'ow - Poland
         \nextinstitute
         ICREA\\
         Lluis Companys, 23 - 08010 - Barcelona - Spain
         \email{reiko.yamada@icfo.eu, eloy.pinol@icfo.eu}
}

\begin{document}

\maketitle

\begin{abstract}

Recently, there has been considerable interest in "sonifying" scientific data; however, sonifying quantum processes using the newest quantum technologies, including Noise Intermediate Scale Quantum devices and quantum random number generators, is still an emerging area of research.

Music technologists and composers employ the growing accessibility to diverse data from quantum mechanics as musical tools in the hope of generating new sound expressions. How different is the quantum world from the classical one, and is it possible to express the quantum world using sounds? Quantum phenomena are very different from those that we experience in our everyday lives. Thus, it is challenging to understand them intuitively. In this paper, we propose sonification as a method toward an intuitive understanding of various quantum mechanical phenomena, from Rabi oscillations and resonance fluorescence of a single atom through the generation of Schr\"odinger cat states in strong laser field physics to insulator-superfluid transition in quantum many-body systems. This paper illustrates various methods we experimented with in sonification and score representations of quantum data depending on the source data and performance settings.

\end{abstract}

\section{Sonification of quantum randomness}\label{sec:1}

When discussing the sonification of data from quantum mechanical systems to sound media, the topic of randomness is unavoidable. We first present our attempt at directly mapping quantum randomness from Rabi Oscillation into musical parameters. We incorporated the data from a quantum random process consisting of a random sequence of spontaneously emitted photons from a single trapped two-level atom, separated by the coherent Rabi oscillations. Such a single quantum trajectory was then used to construct original sound timbres. We have then provided a set of musical timbres originated in quantum randomness to live improvising performers, who then interacted with this "quantum random" input. 

\subsection{Randomness and Music Composition}

Randomness in music composition has a long history. The first famous version of the dice game, {\it Musikalisches W\"urfelspiel} of Mozart was composed and published in 1793 by J.J. Hummel in Berlin-Amsterdam and several different versions were published in the following years \cite{ariza}. Among all the composers who experimented with randomness in composition Iannis Xenakis is probably the composer who greatly influenced the field with his exceptional knowledge and skills in mathematics. In 1962, for example, Xenakis started exploring a stochastic approach to randomness by using computer-based interlinking probability functions to determine composition structure, pitches and their duration \cite{luque}. Generations of composers and music technologists followed his pursuit to explore randomness with various methods of algorithmic compositions. In recent years, composers started to explore randomness using Artificial Intelligence and quantum circuits. In the latter case, the random numbers that they employ most often derive from the true randomness of the genuine quantum origin.

\subsection{Apparent Randomness versus Intrinsic Randomness}

Composers of the 20th and 21st centuries who have experimented with randomness most often used chance operations or random number generators in their creative process. In recent years, algorithmic and stochastic computer-assisted composition and Artificial Intelligence have taken center stage. One of the most notable composers to pave this way is Iannis Xenakis, who, among other methods, used high-speed computations to calculate various probability theories to assist composition processes. The program would "deduce" a score from a "list of note densities and probabilistic weights supplied by the programmer, leaving specific decisions to a random number generator" \cite{alpern}. Nevertheless, in all of these cases, classical random number generators, ergo pseudo-random ones, were used. Thus, the form of randomness used was apparent, which is ultimately deterministic in nature. While their motives for utilizing randomness have been highly diverse, the randomness they employed has almost entirely been apparent randomness. That is to say, the random numbers they employed are imperfect in the strict sense (simply put, perfect random numbers never have repeating patterns) due to the non-genuinely random nature of the process used to generate them. Although each outcome using different systems varies wildly, the difference that using true, intrinsic random numbers would create remains unexplored.

\subsection{Methodology}

In this project, we attempt to take a further step by directly producing sound events from the genuine, true randomness of quantum physical systems. Through this method, we aim to achieve a new aesthetic effect in music, which derives from the true randomness that prevails in the natural quantum world. We experimented with applying random numbers in the reciprocal axis, controlling, for instance, the frequency Fourier spectrum of the sound parameters. The frequency content of sound determines the timbre. Therefore, we explored the possibility of mapping randomness onto timbral changes in sounds. In the quantum world, patterns emerge whenever there is an imbalance in the probability distribution. Consequently, using perfect randomness from quantum physics should prevent the appearance of any patterns in timbral changes, whereas using classical randomness would eventually reveal some.

\begin{figure}[h]
\centering
\includegraphics[width=.45\textwidth]{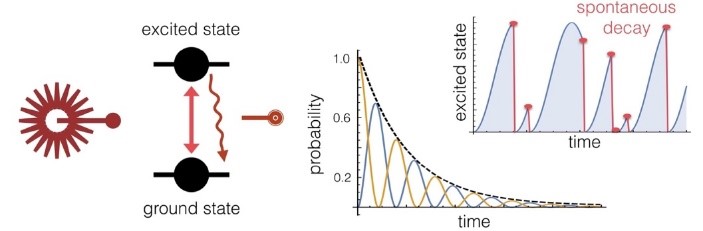}
\caption{Rabi oscillation and spontaneous emission. The inset shows the dynamic of the excited state population.} 
\label{fig1}
\end{figure}

A detailed discussion of the data used for this project can be found in our previous publication \cite{yamada}. In short, we match sinusoidal Rabi oscillations with sound waves containing complex upper harmonics. While the fundamentals of the oscillations always remain intact, the upper harmonics can be divided into several groups corresponding to different timbres. When a transition between two levels of an atom is excited by, say, a laser, its population oscillates between the two states, coherently moving from the lower energy state, the "ground state," and the higher energy one, the "excited state." Depending on the atom's state, when measured, the fundamental is combined with one group of upper harmonics or another, creating a specific sound color in this way. In our approach, we also use another intrinsically random aspect of the studied process: the so-called spontaneous emission. When our atom is excited from the ground state to the excited state, it can spontaneously decay back to the ground state, emitting the "excess" of energy in the form of light (and, more precisely, photons). These events occur at random times, with the waiting time distribution expressing the interplay between the Rabi oscillation and an exponential decay. When mixing the effects of spontaneous emission and Rabi oscillations, the result is simple: the probabilities continue oscillating as they were; then, when the spontaneous emission occurs, the atom falls to the lower energy state, meaning that the probability of finding the atom in this state is one (that is, 100\%). Finally, the probabilities start oscillating again from this point, with a frequency similar to before this event. The new probability for the atom to be excited is shown in Fig. \ref{fig1}. Whenever the atom decays from the excited state to the ground state, a photon is emitted, which can be detected by suitable detectors. Again, the randomness involved in this process stems from the intrinsic random nature of quantum mechanics and is unrelated to an incomplete knowledge of such a system. Note that the rate at which spontaneous emission occurs changes vastly from one system to another, from nanoseconds to even hours in the case of very stable atoms or molecules. The frequency of the Rabi oscillation depends on the physical properties of the atom, as well as on the power of the driving laser light.

 We combine the randomness of the spontaneous decay of the atom with the associated probability of finding it in the excited state. As visible from the inset in Fig. \ref{fig1}, the atom oscillates between the excited and the ground state. However, since the spontaneous decay only happens from a higher to a lower energy state, these events will be more likely to occur when the atom has a higher probability of being in the excited state. Consequently, by collecting the times the atom decays, the resulting histogram would be identical to the probability shown in Fig. \ref{fig1} for a high number of collected events. However, the intermediate histograms, especially the first one, would differ from the final result. If we turn the axis of time into frequency, representing, for example, higher harmonics of a fundamental note and the height of the histogram as intensities, then we could play these harmonics in succession. As spontaneous emission occurs at random times, the time intervals before every timbral change are bound to be genuinely random and obey only the combined probability distributions of spontaneous emission and Rabi oscillation.

\begin{figure}[h]
\centering
\includegraphics[width=.45\textwidth]{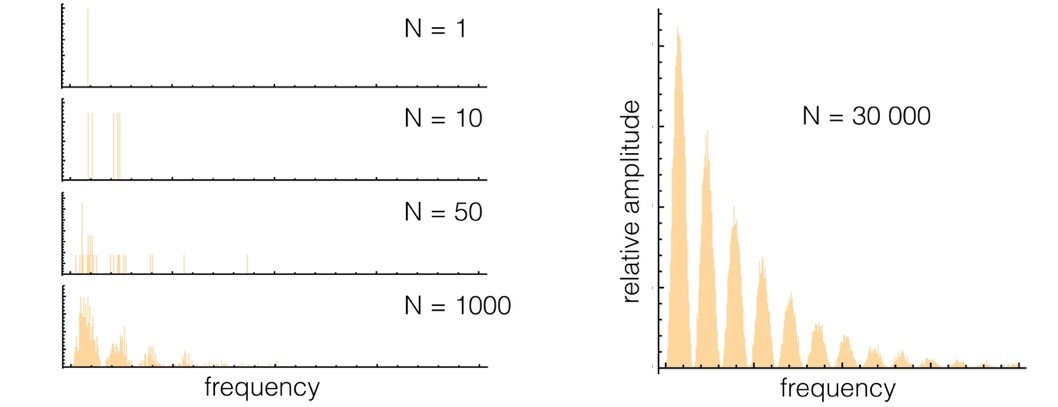}
\caption{Example of histograms of times between consecutive spontaneous emissions, for increasing numbers of simulated events.} 
\label{fig2}
\end{figure}

\subsection{Representation}

The result is sound objects going through changes at random times by the random introduction of an extra harmonic, all according to the probability distribution of spontaneous emission and Rabi oscillations. Such a progression is shown in Fig. \ref{fig2}.

\begin{figure}[h]
\centering
\includegraphics[width=.45\textwidth]{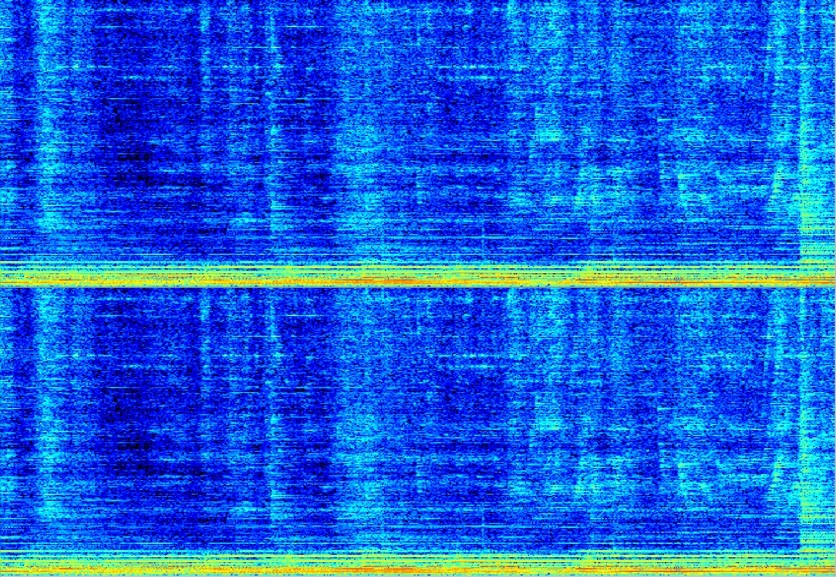}
\caption{Sonogram analysis of a timbre containing multiple fundamentals with rich upper partials based on quantum random data.} 
\label{fig3}
\end{figure}

One of the most original features of our method is that the random events within a specific time frame are represented as one palette of musical timbres (Fig. \ref{fig3}) within one musical event. This method allows the comparison of multiple random samples to be listened to and analyzed in a short amount of time. As a result, we do not convert certain random phenomena into numbers and then convert them again into sound parameters. In contrast, the intrinsic random phenomena here are directly seen and heard through timbres one by one, a few seconds apart.

\begin{figure}[h]
\centering
\includegraphics[width=.45\textwidth]{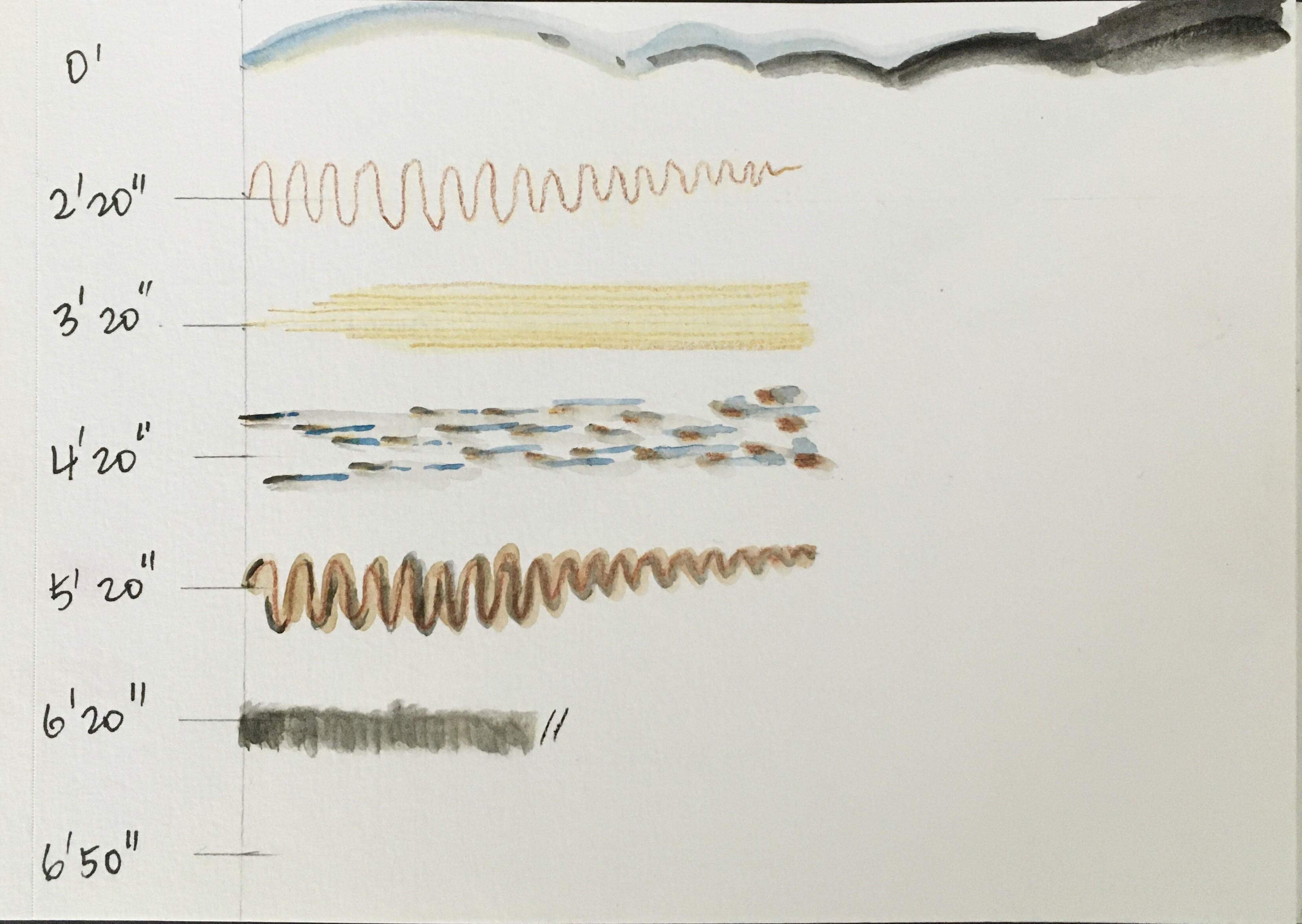}
\caption{Example of graphic notation.} 
\label{fig4}
\end{figure}

\begin{figure}[h]
\centering
\includegraphics[width=.45\textwidth]{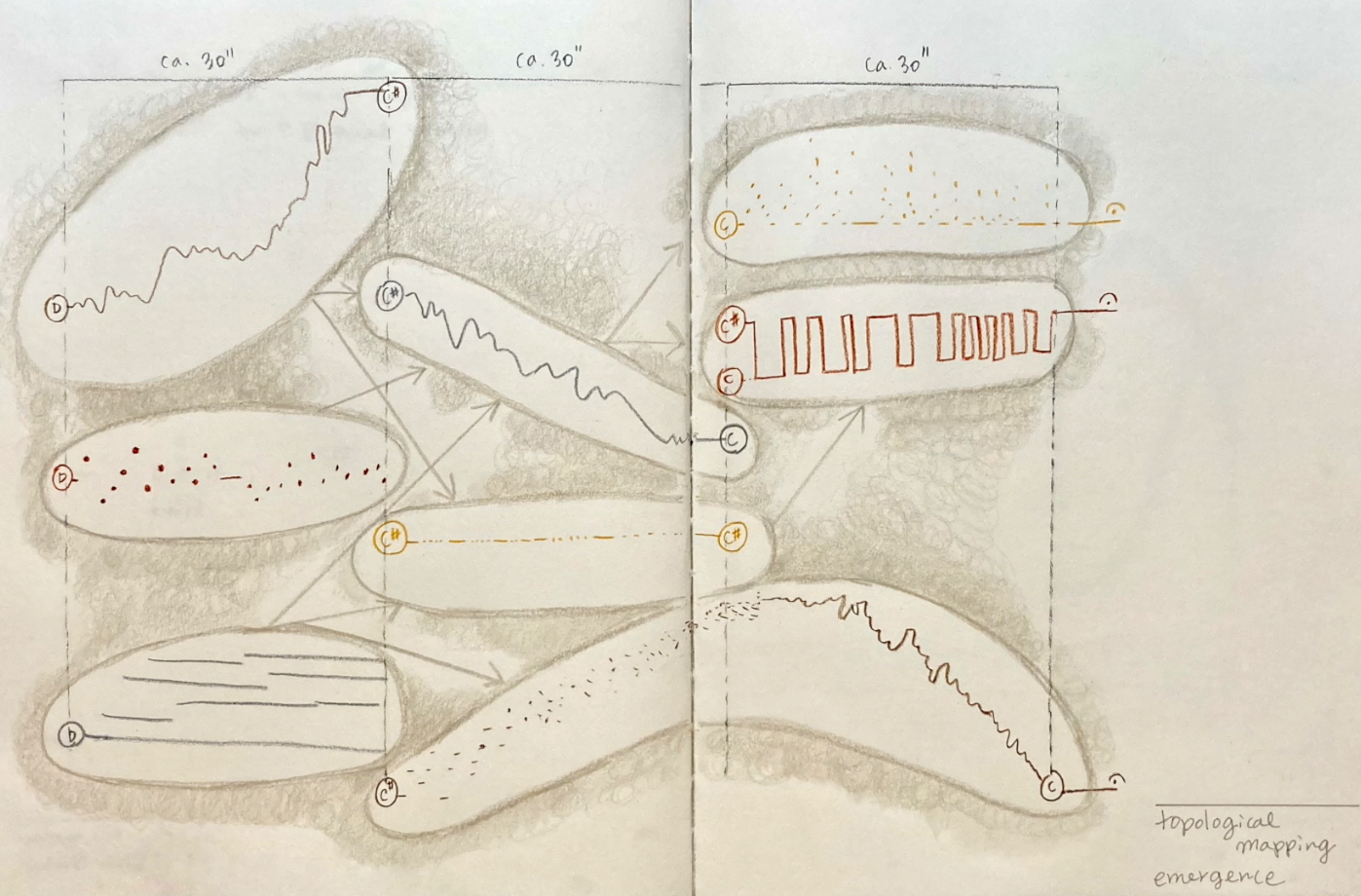}
\caption{Example of moving image score.} 
\label{fig5}
\end{figure}

\begin{figure}[h]
\centering
\includegraphics[width=.45\textwidth]{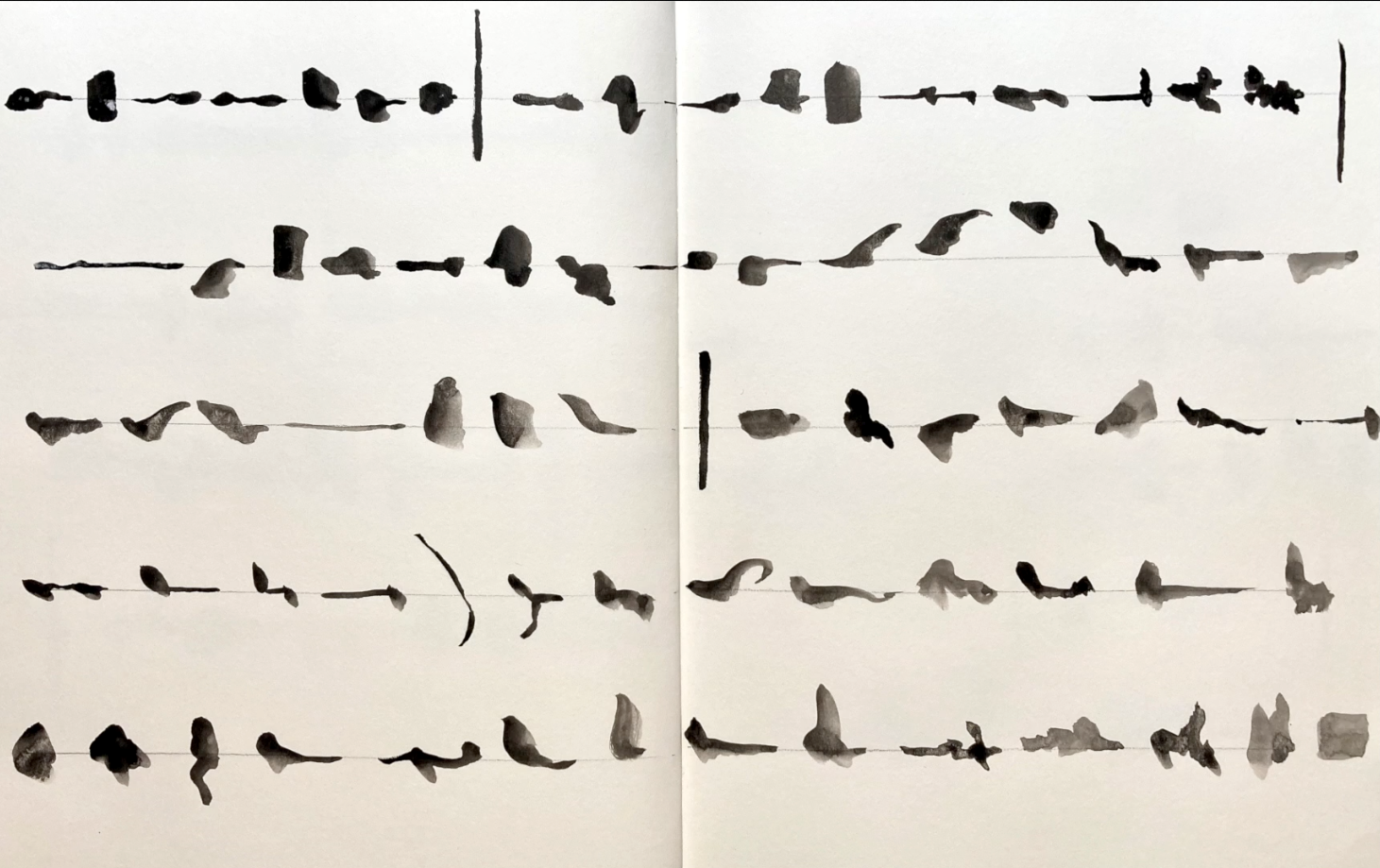}
\caption{Example of moving image score.} 
\label{fig6}
\end{figure}

\begin{figure}[h]
\centering
\includegraphics[width=.45\textwidth]{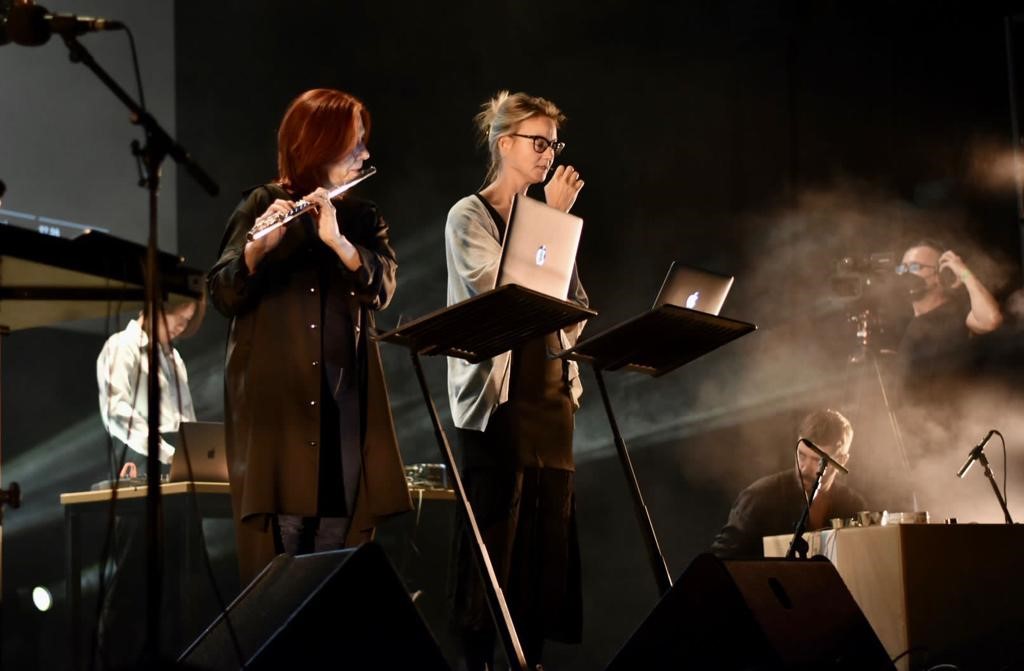}
\caption{Musicians following the moving image score. Our approach was used in the concert "Interpreting Quantum Randomness" at Sonar Festival in Barcelona 2021.} 
\label{fig7}
\end{figure}

In order to guide free improvising musicians to follow the series of newly created timbral events, we created a visual guide to serve as a simple musical score. The score is further categorized into two types: graphic notation (cf. Fig. \ref{fig4}) and moving image (cf. Fig. \ref{fig5} and Fig. \ref{fig6}). The moving image helped the performers to follow the sequence of timbral sonority that originates in quantum random data (see Fig. \ref{fig7}).

\section{Sonification of the Wigner function}

As a second example, we demonstrate how we attempted to sonify the quantum Wigner function for acousmatic composition and string quartet score using several methods.

\subsection{Wigner function}

In quantum physics, the position $x$ and momentum $p$ (mass times velocity) of a particle have a different nature than in classical physics. Mathematically, they are described by objects called operators (matrices), which do not have to commute one with another, meaning that $xp$ is not the same as $px$. This fact has profound physical consequences: In quantum mechanics, a particle cannot have a determined position and momentum simultaneously. If the particle is localized with standard deviation $\Delta x$ in position and with standard deviation $\Delta p$ in momentum, then according to the famous Heisenberg's uncertainty principle, one obtains the inequality: $\Delta x \Delta p \ge \hbar$, where $\hbar$ is Planck's constant, divided by 2$\pi$. In effect, if one wants to localize the particle, ideally with $\Delta x$ tending to zero, $\Delta p$ must tend to infinity, and vice versa \cite{braginsky}. The so-called quadratures of the quantized oscillating electric field, which classically are parts of the electric field that oscillate like {\it sine} or {\it cosine}, respectively, have similar properties. 

A further consequence of this is that, in contrast to classical mechanics, one cannot define the joint probability distribution of $x$ and $p$ in a mathematically sound way. What one can do, however, is to define a so-called quasi-probability distribution, called Wigner function $W(x,p)$, which takes real number values but may admit negative values and necessarily does for "massively quantum" states of the particle \cite{schleich}. In this sense, the Wigner function differs significantly from a classical probability distribution. However, it has many appealing properties: i) it is defined in such a way that integrated over $x$ and $p$ gives 1, i.e., is normalized like a standard probability distribution; ii) integrated over $x$ ($p$) gives a well-defined probability distribution of $p$ ($x$). The Wigner function is a handy mathematical object to represent complete information about quantum states, in particular, intrinsic and correlated randomness of the position and momentum - that is why we use it in our sonification protocol.

\subsubsection{Data used}

The data comes from the Wigner quasi-probability distribution description of two different types of quantum states: Fock and Schr\"odinger cat states. Fock states are characterized by having a well-defined number of particles $m$. This makes them very special as for "typical" quantum mechanical states, such as coherent or squeezed states, the number of particles is not fixed and is not well defined; this leads to different results when we measure it, expressed by the corresponding particle-counting distribution. The following expression describes the Wigner function of the $m$th Fock state:
\begin{equation}
W_m (x,p)=\frac{(-1)^m}{\pi} e^{-(x^2+p^2 )} {\cal L}_m \left(2(x^2 + p^2)\right),
\end{equation}
where ${\cal L}_m$ denotes the $m$th Laguerre polynomial. 

\begin{figure}[h]
\centering
\includegraphics[width=.45\textwidth]{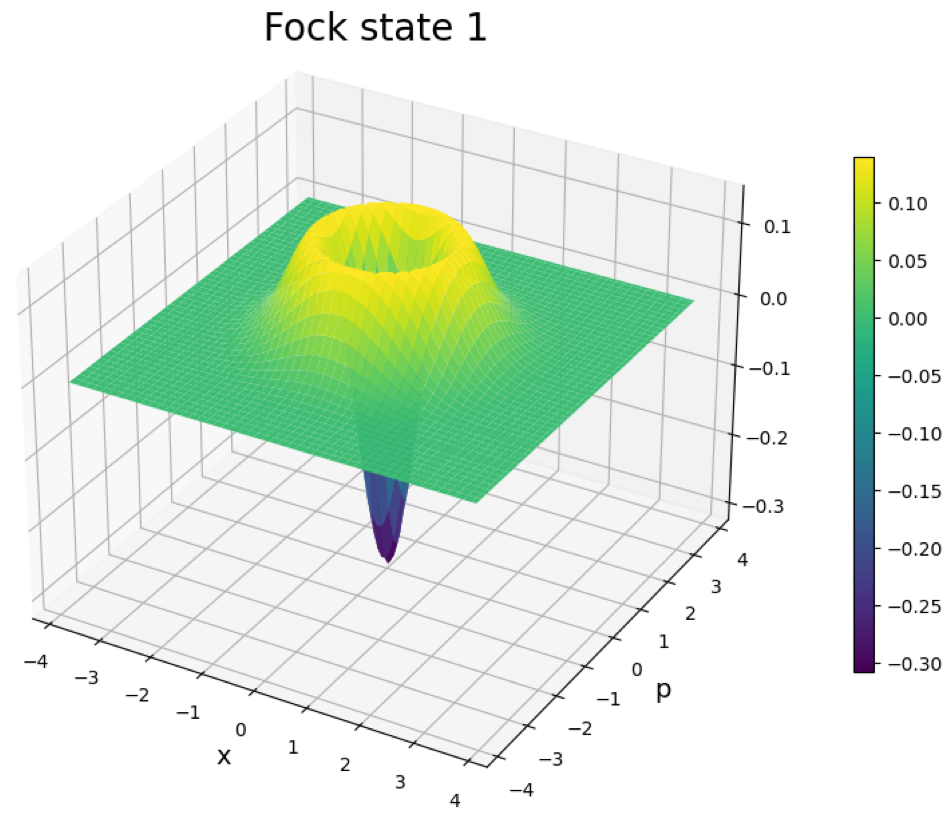}
\caption{Wigner function of the Fock state $m=1$.} 
\label{fig8}
\end{figure}

\begin{figure}[h]
\centering
\includegraphics[width=.45\textwidth]{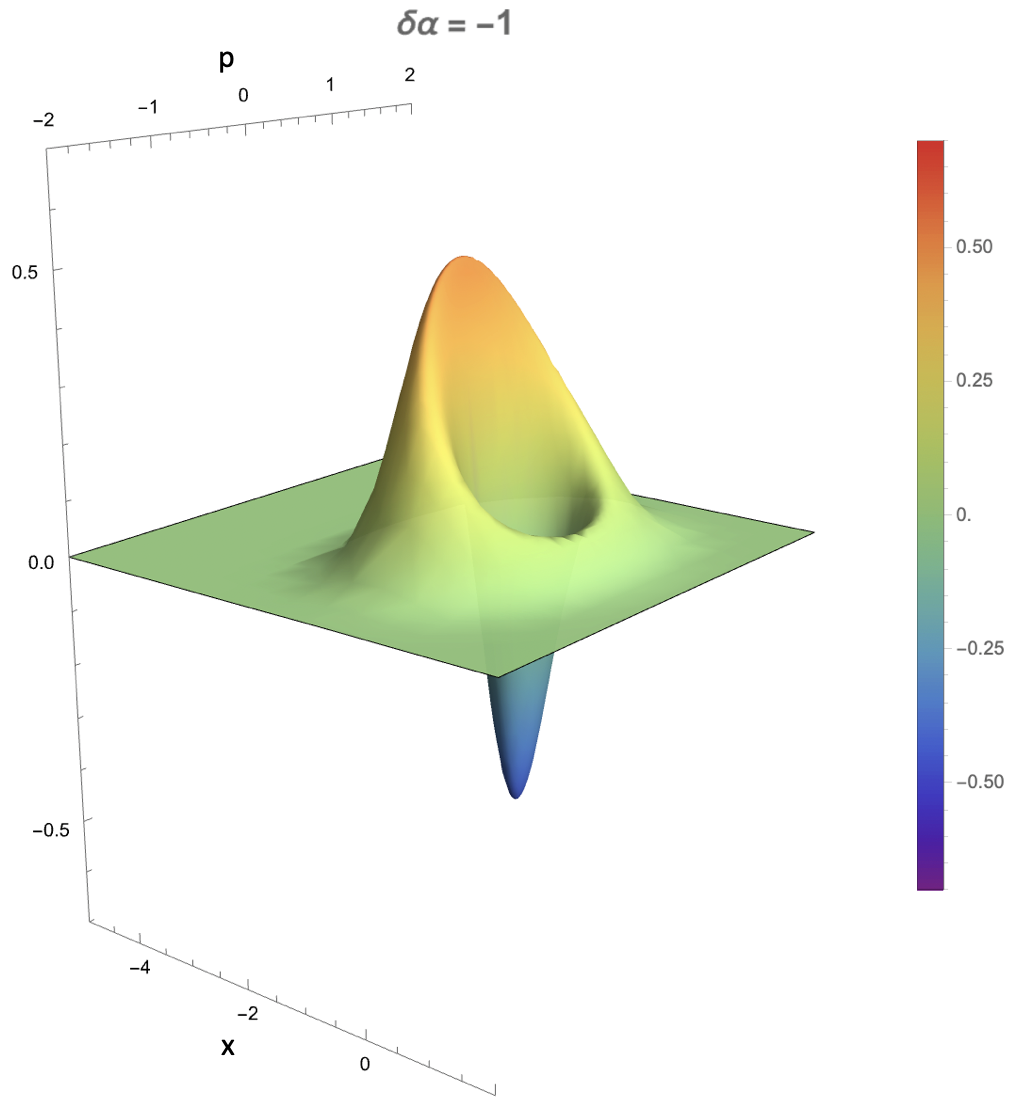}
\caption{Wigner function of the optical kitten state with $\delta\alpha=-1$.} 
\label{fig9}
\end{figure}

Schr\"odinger cat states are in general quantum superposition of at least two different macroscopic (or mesoscopic) states (such as "cat alive" and "cat dead"). The latter refers to the thought-experiment proposed by Erwin Schr\"odinger in 1935, where the cat is in a superposition of alive and dead states. In our case, we used a particular kind of cat states, namely optical cat states, or better to say, kitten states. These states, superpositions of the macroscopically different high-photon-number coherent-states, have been recently predicted and observed in high-harmonic generation and above-threshold ionization processes in Xenon atoms \cite{lewenstein, stammer}. Their Wigner function has the form:
\begin{equation}
\begin{split}
W(\beta)= & \, \frac{2}{\pi(1-e^{-|\delta\alpha|^2 })}\bigl[e^{-2|\beta-\alpha-\delta\alpha|^2 } \\
&+e^{-|\delta\alpha|^2 } e^{-2|\beta-\alpha|^2 } 
-e^{-|\delta\alpha|^2}e^{-2|\beta-\alpha|^2} \\
&\times (e^{2(\beta-\alpha)\delta\alpha^* }+e^{2(\beta-\alpha)^*\delta\alpha})\bigr].
\end{split}
\end{equation}
Here, $\alpha$ and $\beta$ are complex numbers related to the "classical" arguments of the Wigner function via $\beta -\alpha = x+ip$. We can see an example of both Fock and cat states in Figures \ref{fig8} and \ref{fig9}. Depending on the value of $\delta \alpha$, the shift between the two superposed states, the final state can be in a cat or a coherent state. Then, we can simulate a transition from a quantum state to an almost classical one. In some of the mapping methods that we show, we have discretized the arguments of the Wigner functions and their values, making a grid of points in order to make our analysis more accessible. The grid limits are chosen to contain at least 99\% of the quasi-probability distribution. We discretized in two different ways: Making the grid regular (all the points are equidistant) or making the grid non-regular, using intervals extracted from a Gaussian probability distribution. In the second case, the mean and the standard deviation of the distribution correspond to the average length of the intervals in the regular case and to half of it, respectively. The latter is chosen in this way because the ranges of {\it values} $[\mu-\sigma,\mu +\sigma]$ and $[\mu-2\sigma,\mu +2\sigma]$ contain approximately 68.26\% and 95.44\% of the probabilities to be given. This means we do not have to deal with extensive intervals.

\subsubsection{Mapping method}

We have used different methods to map the quantum states to sounds: 
\begin{enumerate}
    \item [(a)] Mapping each point of the grid $(x, p, W)$ to phase, frequency, and amplitude, respectively, of different types of waves, as sinusoidal or triangular ones, on the audible range. Then, we reproduced all of them together using SuperCollider \cite{super}, an environment and programming language for real-time audio synthesis and algorithmic composition. Due to hardware limitations, we could only play up to approximately 900 sound waves simultaneously. Hence, we worked on the cases where the grid had 10, 20, and 30 points per side (i.e., 100, 400, and 900 points on the regular grid). 
    \item [(b)] Another method is to acquire the minimum and maximum values of the Wigner function and map them to frequencies.
    \item [(c)] Another alternative method is to cut the whole figure horizontally into four different parts with the same height and analyze their WF values at half height and their volumes. The latter is related to the quantum behavior of the state and is mapped to the intensity of each frequency or tremolos type.
    \item [(d)] Finally, we have also mapped properties that are hard to recognize visually, such as the moments of the distribution.
\end{enumerate}

Note that with those methods, each state produces a "stationary" sound. Then, if the state changes, for example, varying the value of $\delta \alpha$, the sound also changes. The mappings were done through linear and quadratic functions obtained with regressions. The regressions for the mappings of (a) are linear. Their minimum and maximum values are 0 and 2$\pi$ in the case of $x$ (0 and 4 when we map them to triangular waves) and 440 Hz and 1760 Hz in the case of $p$. They correspond to the minimum and maximum values of $x$ and $p$,  which we used in the discretized grid.
On the other hand, we have also mapped the values of the Wigner function directly to the amplitudes of the waves. The Wigner function is normalized (its volume equals 1), but here, we are summing their values, not the volumes, so that it can be very different from 1. We attempted two different combinations of waves: in both cases, the points corresponding to positive Wigner function values were mapped to sinusoidal waves, but in one case, the points corresponding to negative Wigner function values were mapped to triangular waves. In another case, we mapped them to sinusoidal waves with a pulsed wave controlling them with an arbitrary frequency (we used 0.5 Hz).

On the other hand, in (b) and (c), we have used quadratic functions to map the values of the Wigner function to frequencies. In these cases, the regression was done in a way that the minimum and maximum values of the WF in the range $\delta\alpha \in (0, -3)$ are mapped to 146.83 Hz and 1318.5 Hz, respectively, and the 0 value to 466.16 Hz. In (c), the mapping of the volumes of each chunk to the intensity of the sound is made in a similar way as the frequency one. We looked for the largest positive and negative values of the volumes of the four different chunks. Subsequently, we created two intervals of the same length between 0 and the largest (absolute) value of the negative volumes and six intervals between 0 and the largest value of the positive volumes. Then, we check in which interval the volume of each chunk fits and assign a number between -2 and 6 (without 0) to it, which will represent its intensity. Note that the negative intensities correspond to negative volumes, which are a quantum feature.

Finally, in (d), we have mapped the first and second moments of $x$, corresponding to the mean and the standard deviation of the distribution with respect to that variable, to the first and second moments of a Gaussian sound produced with SuperCollider. By Gaussian sound, we refer to a sound that, analyzed in a frequency-intensity graph, has a Gaussian profile. The mapping is done in a way related to the frequency distribution of the keys of a piano in twelve-tone equal temperament, with the 49th key tuned to 440 Hz.
Using the abovementioned methods, we mapped the data into both fixed media and traditional Western notation for acoustic instruments (string quartet). 

\subsubsection{Instrumentation}
For fixed media, all frequencies were mapped with neutral sinusoidal waves as well as triangular pulses, with the phase and intensity of each frequency respectively controlled by the Wigner function data.

We also explored the possibilities for mapping variables for string instruments. We used the traditional Western notation in this case.

\subsubsection{Score representation}

\begin{figure}[h]
\centering
\includegraphics[width=.45\textwidth]{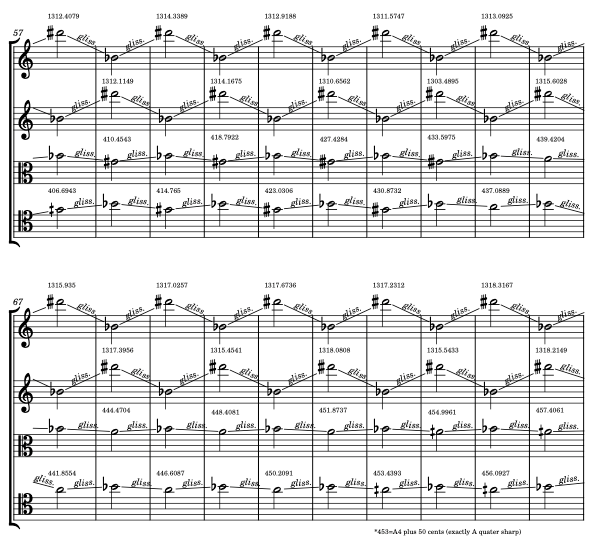}
\caption{Score example of Wigner function representation in String Quartet obtained with the mapping method (b).} 
\label{fig10}
\end{figure}

\begin{figure}[h]
\centering
\includegraphics[width=.45\textwidth]{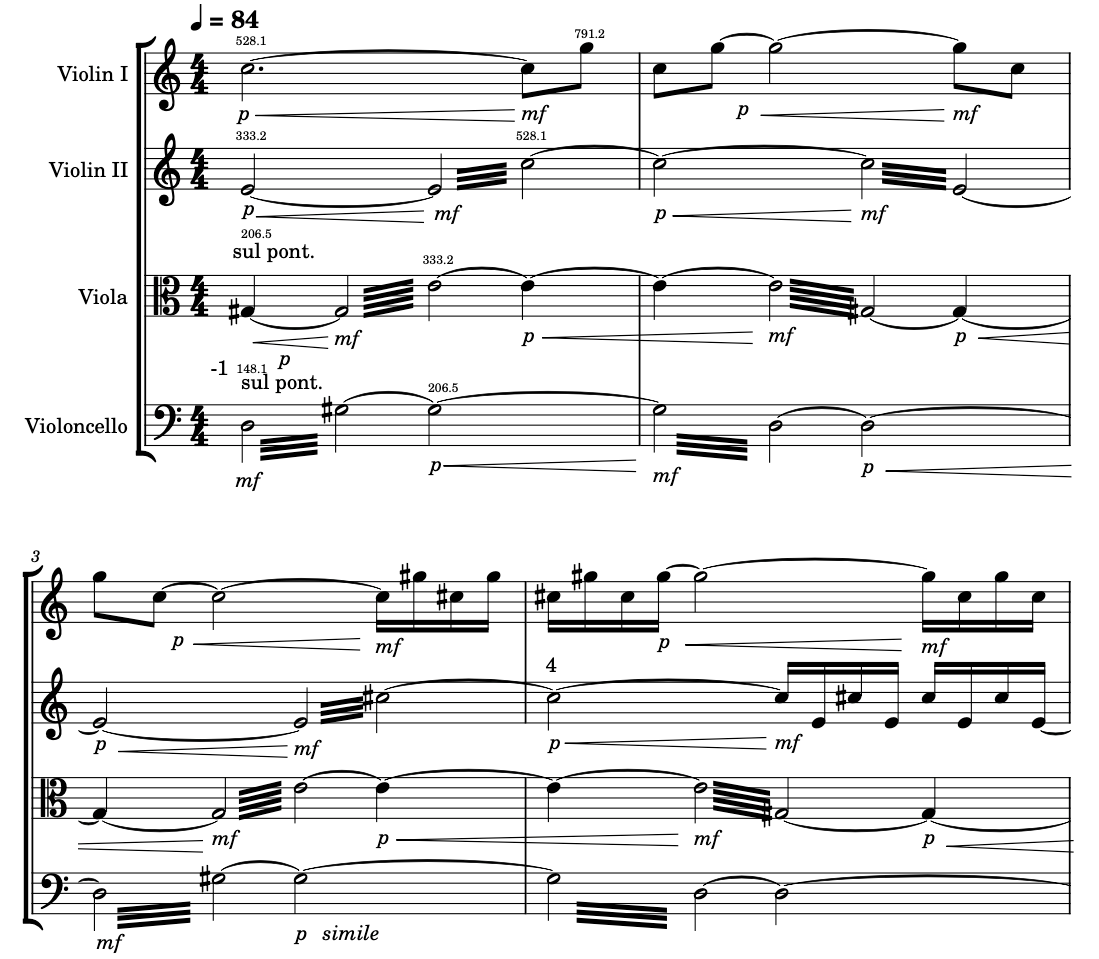}
\caption{Score example of Wigner function representation in String Quartet obtained with the mapping method (c).} 
\label{fig11}
\end{figure}

In the mapping methods (b) and (c), all frequencies were rounded up to the nearest quarter divisions of the 12-note equal tempered notes. Still, exact frequencies from the mapping are indicated above each note in case performers decide to make further divisions of notes as seen on the score example Figure \ref{fig10} (for the corresponding hyperlink to the sound file, see \cite{fig10}). In (b), the pitches coming from the maximum values of the WF are played by the two violins. In contrast, the ones coming from the minimum values are played by the viola and the violoncello. In the mapping method (c), we separated the WF into four different vertical sections. Each section was mapped to each instrument in the string quartet and scored accordingly. In order to cover the entire WF values with frequencies among four instruments, we experimented with glissandi (Figure \ref{fig10}) and tremolos (Figure \ref{fig11}; for the corresponding hyperlink to the sound file, see \cite{fig11}). The negativity of the WF is represented through different playing techniques, such as the sul ponticello or the ricochet bowing ones.

We explored the option of mapping the volumes of the four parts into the intensity of sounds (noted in dynamics markings) or tremolo types. The latter method represents better the quantum state of the Wigner function discussed above. Finally, in the last presented method (d), the musicians do not need a score as they are free to play any single pitch contained in the Gaussian sound as an analogy to the collapse of quantum states when measured.

\section{Sonifying Quantum Simulations}

In this last section we propose to sonify the Mott insulator-superfluid quantum phase transition in the Bose-Hubbard model. We will use the numerically simulated data and experimental results in the future.

Actual quantum processors, often referred to as NISQ (Near Intermediate-Scale Quantum), encounter susceptibility to errors arising from qubit decoherence and environmental noise, presenting significant challenges in quantum error correction. The concept of quantum simulation is safer since it does not rely on quantum error correction. As discussed in \cite{lnp1000}, there are already many successful platforms and architectures and new challenges for the coming decades. 

\subsection{Quantum Simulators}

A \textit{simulator} is a device that emulates the desired properties of a given system, such as an airplane, train, or a physical system of quantum particles. Airplane simulators cannot fly nor take passengers on board. However, it provides a realistic reproduction of all navigational, mechanical, and electronic apparatus necessary for efficient and safe crew training. Similarly, quantum simulators (QS) do not reproduce all properties of the simulated systems, as a perfect general quantum computer would do, but imitate only the specific features required to understand the phenomena of interest occurring in the original system. For the interest of the reader, refer to \cite{Miranda2022} to see some examples and ideas of how current quantum computers can be used for music composition. The general idea and the concept of such QS can be shortly sketched as follows: 

\begin{itemize}
\item There exist many interesting {\bf quantum phenomena} with highly important applications (such, for instance, superconductivity).
\item These phenomena are often complex to describe and understand with the help of standard or even supercomputers.
\item Maybe we can design another, simpler, and more controllable quantum system to simulate, understand, and manipulate these phenomena, as proposed originally by Yu. I. Manin and R. P. Feynman. Such a system would thus work as a quantum computer of one specific purpose, i.e., a {\bf quantum simulator}.
\end{itemize}

Designing such a simulator is, however, a challenging task. The research on QS goes back to the beginning of this century, and there are numerous valuable reviews covering the various platforms and types of QSs. In fact, the beginning of the practical concept of QS goes back to the proposal for simulating strongly correlated systems in optical lattices and the first experiments. Nowadays, QS is commonly used for the following tasks and goals:

\begin{itemize}
\item {\bf Fundamental problems of physics.} This is the most developed application, in which many achieved results are believed to reach a quantum advantage; this is particularly true for the studies of quantum dynamics or disordered quantum systems, such as the ones that exhibit many-body localization (MBL).
\item {\bf Quantum chemistry.} Applications of quantum NISQ devices and QS to quantum chemistry have only started. Although promising, it is still far from achieving the precision and accuracy of contemporary theoretical quantum chemistry.
There is growing evidence that the expectation of an exponential quantum advantage in quantum simulations for quantum chemistry is unrealistic.
\item {\bf Classical/quantum optimization problems for technology.} Applications of quantum NISQ devices and QS to optimization problems are also in the initial phase and cannot yet compete with the classical supercomputer methods.
\end{itemize}

\subsection{Bose Hubbard model and Mott insulator -- superfluid transition}

\begin{figure}[t]
\centering
\includegraphics[width=6cm]{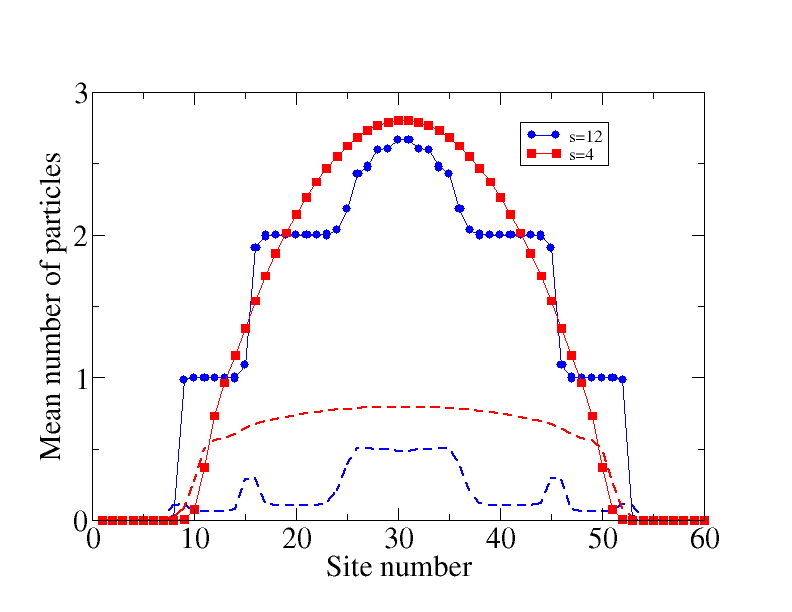}
\caption{Exemplary results for 1D Bose-Hubbard model: Population of 80 atoms distributed in 60 sites, plotted for various lattice depths $s$ in the units of recoil energy. Broken lines describe the standard deviation of populations. For small $s$ the system is superfluid, described by a smooth distribution of populations, but with a large standard deviation. For larger $s$ the system enters the Mott insulating regime and  the populations exhibit "wedding cake"  structures with smaller standard deviation except for tiny regions between different population levels where SF is squeezed in.} 
\label{fig12}
\end{figure}

One of the most paradigmatic applications of QS is to simulate quantum phases transition in the Bose-Hubbard (BH) model (for an overview of the quantum simulators, see \cite{LSA2017}). BH model describes a system of bosonic atoms in an optical lattice  that can tunnel from one site to other, and interact pairwise repulsively. Another relevant energy scale is set by the chemical potential $\mu$, which controls the number of atoms in the system. When tunneling energy $t$ dominates over the interaction energy $U$, the system is in the superfluid (SF) phase, characterized by the long range phase coherence of the bosonic wave function and fluctuations of density. In contrast, when interactions dominate, the systems enters a Mott insulator (MI) phase, where  the density fluctuations are reduced, and ideally we have one (two, three...) atoms per site. In practice, experiments are done in a shalow harmonic potential which acts like a  local chemical potential, creating "wedding cake" structure with say MI  with 2 atoms per site in the middle of the trap, a ring of SF, a ring of MI with one atom per site, and another ring of SF. This is illustrated very well in Fig. \ref{fig12}, where the 1D Bose-Hubbard models was investigated.

\subsection{Methods} 
The idea is to consider a 2D system on a 20x20 lattice with bosons in a  loose trap and follow them, changing the parameters of the system going from insulator to superfluid phase. The data should correspond to the occupations of each site. In the Mott insulator phase, they should be more or less one or two atoms per site, with the wedding cake structure. In the superfluid,  there will be a smooth density profile but with significant fluctuations.

Since this will correspond to the data set of 20x20 (= 400) describing the number of atoms in each site, it can be sonified as done with Wigner Function. We start by setting the system deeply in the MI state and vary parameters to enter the SF phase. We fully control how fast parameters vary to achieve the phase transition from MI to SF. At each time instance, we get 400 + 400 numbers describing mean populations and their standard deviations at the lattice sites. Such data not only can be generated numerically but is also accessible experimentally (see below). 

\subsection{Single site and single particle resolution}
 
A quantum gas microscope is capable of detecting single atoms in a single site of the (optical) lattice. In the first experiments of the Harvard and Garching groups,  the device could only distinguish between zero/two or one bosonic atoms at the site. Still, it offered an unprecedented possibility of measuring density-density correlations far beyond the possibilities of standard condensed matter physics. These devices are nowadays capable of measuring single particles in a single site of the lattice with spin resolution for both bosons and fermions and even more (for a review \cite{Kuhr}).


\section{Conclusions and Outlook}

We demonstrated various solutions to the {\it sonification} and the representation of quantum phenomena, each requiring a unique approach and solution depending both on the data of their quantum origin and the instrumentation of the artistic outcome. Dealing with live musicians required further mapping into the score in the traditional Western or graphic notation (fixed or moving image). In each case, we attempted to select the method likely to capture an intuitive representation of each quantum phenomenon.

Creating music composition that provides an intuitive understanding of quantum physics is highly challenging, and we are still at the beginning of its experimentation. Still, among all the fascinating and intricate quantum phenomena, randomness is an essential topic closely related to music composition, and our experimentation of {\it sonifying} quantum randomness showed promising results. We will continue to explore various {\it sonification} methods of quantum phenomena, using a similar approach to our method of sonifying quantum randomness. For example, we are exploring methods of mapping wave function values to audio filtering. The method used till now resembles subtractive synthesis; however, it requires some approach to calculate yet another dimension (axis). An early experiment suggests that this method might represent the time transformation of the Wigner Function better than other methods.

Finally, in the last section of this work,  we speculated about the sonification of Quantum Simulation, trying to answer the question, "Can we hear superfluidity? Can we hear superconductivity?" If we are successful in this endeavor, we will be a step closer to fully expressing the quantum world through sonification.


\noindent{\bf Acknowledgments}

The academic and artistic research as well as performance mentioned in this article were made possible by ICFO, Phonos Foundation, Sonar Festival and Phonos Foundation (Barcelona, Spain). Figures 1 and 2, and data associated to Rabi oscillations are provided by Samuele Grandi (ICFO). The authors would like to thank Philipp Stammer (ICFO) for his generous help with topics related to Wigner function. 

R.Y, E.P. and M.L. acknowledge support from: ERC AdG NOQIA; MICIN/AEI (PGC2018-0910.13039/501100011033, CEX2019-000910-S/10.13039/501100011033, Plan National FIDEUA PID2019-106901GB-I00, FPI; MICIIN with funding from European Union NextGenerationEU (PRTR-C17.I1): QUANTERA MAQS PCI2019-111828-2); MCIN/AEI/ 10.13039/501100011033 and by the “European Union NextGeneration EU/PRTR"  QUANTERA DYNAMITE PCI2022-132919 within the QuantERA II Programme that has received funding from the European Union’s Horizon 2020 research and innovation programme under Grant Agreement No 101017733Proyectos de I+D+I “Retos Colaboración” QUSPIN RTC2019-007196-7); Fundació Cellex; Fundació Mir-Puig; Generalitat de Catalunya (European Social Fund FEDER and CERCA program, AGAUR Grant No. 2021 SGR 01452, QuantumCAT \ U16-011424, co-funded by ERDF Operational Program of Catalonia 2014-2020); Barcelona Supercomputing Center MareNostrum (FI-2023-1-0013); EU (PASQuanS2.1, 101113690); EU Horizon 2020 FET-OPEN OPTOlogic (Grant No 899794); EU Horizon Europe Program (Grant Agreement 101080086 — NeQST), National Science Centre, Poland (Symfonia Grant No. 2016/20/W/ST4/00314); ICFO Internal “QuantumGaudi” project; European Union’s Horizon 2020 research and innovation program under the Marie-Skłodowska-Curie grant agreement No 101029393 (STREDCH) and No 847648  (“La Caixa” Junior Leaders fellowships ID100010434: LCF/BQ/PI19/11690013, LCF/BQ/PI20/11760031,  LCF/BQ/PR20/11770012, LCF/BQ/PR21/11840013). E.P. is supported by ``Ayuda (PRE2021-098926) financiada por MCIN/AEI/ 10.13039/501100011033 y por el FSE+". The work of J.Z. was funded by the National Science Centre, Poland, project 2021/03/Y/ST2/00186 within the QuantERA II Programme that has received funding from the European Union Horizon 2020 research and innovation programme under Grant agreement No 101017733.

\bibliographystyle{unsrt}
\bibliography{ms.bib}

\end{document}